\documentclass[twocolumn,a4paper]{revtex4}

\usepackage{amssymb, amsmath,framed}
\usepackage[colorlinks]{hyperref}
\usepackage[english]{babel}
\usepackage{graphicx}
\usepackage{graphics}
\usepackage{subfigure}
\usepackage{epstopdf}
\usepackage{epsfig}

\newcommand{\rem}[1]{}
\newcommand{\de}{{\rm d}}

\newcommand{\bx}{{\mathbf{x}}}
\newcommand{\bq}{{\mathbf{x}}}
\newcommand{\bv}{{\mathbf{v}}}
\newcommand{\bk}{{\mathbf{k}}}
\newcommand{\bp}{{\mathbf{p}}}
\newcommand{\bm}{{\mathbf{m}}}
\newcommand{\bM}{{\mathbf{M}}}

\newcommand{\bK}{{\mathbf{K}}}
\newcommand{\bE}{{\mathbf{E}}}
\newcommand{\bB}{{\mathbf{B}}}

\newcommand{\bu}{{\boldsymbol{u}}}
\newcommand{\bU}{{\boldsymbol{U}}}
\newcommand{\bV}{{\boldsymbol{V}}}
\newcommand{\bW}{{\boldsymbol{W}}}
\newcommand{\dvol}{{\de^3\bx\,\de^3\bp}}
\newcommand{\tU}{{{\widetilde{\boldsymbol{U}}_1}}}
\newcommand{\tK}{{\widetilde{\mathbf{K}}_1}}

\newcommand{\tp}{{\widetilde{\mathsf{p}}_1}}

\newcommand{\tf}{{\widetilde{f}_1}}
\newcommand{\tB}{{\widetilde{\mathbf{B}}_1}}
\newcommand{\bX}{{\mathbf{X}}}
\newcommand{\z}{{\mathbf{e}_z}}

\newcommand{\beq}{\begin{equation}}
\newcommand{\eeq}{\end{equation}}
\newcommand{\ben}{\begin{eqnarray}}
\newcommand{\een}{\end{eqnarray}}

\def\Bbb{\mathbb}
\newcommand{\bJ}{{\mathbf{J}}}

\begin{document}

\title{Hybrid Vlasov-MHD models: Hamiltonian vs.\  non-Hamiltonian}
\author{Cesare Tronci}
\email{c.tronci@surrey.ac.uk}
\affiliation{Department of Mathematics, University of Surrey, Guildford GU2 7XH, United Kingdom}
\author{Emanuele Tassi}
\email{emanuele.tassi@cpt.univ-mrs.fr}
\affiliation{CNRS \& Centre de Physique Th\'eorique, Campus de Luminy, 
13288 Marseille cedex 9, France\\and Universit\'e de Toulon, CNRS, CPT, UMR 7332, 83957, La Garde, France}
\author{Enrico Camporeale}
\email{e.camporeale@cwi.nl}
\affiliation{Centrum Wiskunde \& Informatica, 1098 XG Amsterdam, Netherlands}
\author{Philip J. Morrison}
\email{morrison@physics.utexas.edu}
\affiliation{Department of Physics \& Institute for Fusion Studies, University of Texas, Austin 78712-0262, USA}

\begin{abstract}

\bigskip

This paper investigates hybrid kinetic-MHD models, where a hot plasma (governed by a kinetic theory) interacts with a fluid bulk (governed by MHD).  Different nonlinear coupling schemes are reviewed, including the pressure-coupling scheme (PCS) used in modern hybrid simulations. This latter scheme suffers from being non-Hamiltonian and to not exactly conserve  total energy.   Upon adopting the Vlasov description for the hot component,   the non-Hamiltonian PCS and a  Hamiltonian variant are compared. Special emphasis is given to the  linear stability of Alfv\'en waves, for  which  it is shown that a  spurious instability appears at high frequency in the non-Hamiltonian version. {This instability} is removed in the Hamiltonian version. 
$\,$\\
\bigskip
\end{abstract}

\maketitle



\section{Introduction}
\label{intro}


Several configurations in plasma physics involve the interaction of a hot plasma species with a lower temperature bulk component. Typical {examples}  are those of nuclear fusion devices, in which the energetic alpha particles produced by the fusion reactions interact with the ambient plasma, and those {of space plasmas, involving}  the interaction between the energetic solar wind and Earth's magnetosphere.   Such plasmas  have been studied for decades and yet continue to be the subject of current research.

For such configurations, one is first interested in ascertaining the stabilizing or destabilizing effects that the energetic component can have on the overall system. In order to address this question, various mathematical models have been  formulated to include the combined effects of both the energetic particles and the bulk plasma.  Although the bulk can be well described by ordinary magnetohydrodynamics (MHD),  adequately modeling the hot species requires the use of kinetic theory.  This  multiscale, multi-physics approach leads to the formulation of hybrid kinetic-MHD models that  couple the MHD equations to a kinetic equation for the hot component.   Then,  the question of which kinetic equation  to use for the hot particles arises.  Typically drift-kinetic, gyrokinetic,  or the  full Vlasov system are used.  In plasma fusion, the first two options are used most often, while the full Vlasov description is needed for \emph{e.g.} reverse field pinch plasmas \cite{Kim}. The full Vlasov description solves for effects at all scales and thus is less convenient when the hot particle gyromotion can be averaged out, in favor of drift-kinetic and gyro-kinetic models. Nevertheless, this paper aims to account for hot particle effects at all possible scales, so that the full Vlasov description is adopted.

Another more important question emerges  in the formulation of hybrid kinetic-MHD models, viz.,  the particular type of coupling scheme that should be used in the model. Two coupling schemes are present in the literature: the current-coupling scheme (CCS), found for example in Refs.~\cite{BeDeCh,ChWhFuNa,ToSaWaWaHo} and the pressure coupling scheme (PCS), examples of which are used in Refs.~\cite{Cheng,FuPark,KimEtAl}.  While the CCS involves the hot {momentum} and density
\ben
{
\bK}&=&\!{\int\!\bp\,f(\bx,\bp)\,\de^3\bp}\,,
\\
 n&=&\!\int\!f(\bx,\bp)\,\de^3\bp\,,
\een
the PCS involves the  following  tensor: 
\begin{equation}
\Bbb{P}=\frac1{m_h}\int\!\bp\bp\,f(\bx,\bp)\,\de^3\bp\,,
\label{presstensor}
\end{equation}
which is the pressure tensor of the  hot component, calculated with respect to a zero mean velocity.  All these quantities are defined as above in terms of moments of the kinetic probability density $f(\bx,\bp)$ on phase space. Here, $\bp$ denotes the kinetic momentum $\bp=m_h\mathbf{v}$, while $m_h$ and $q_h$ denote the hot particle mass and charge, respectively. Normally, the PCS is derived from the CCS \cite{PaBeFuTaStSu,ParkEtAl}, under the assumption that the hot component is rarefied, so that its density levels are much lower than those of the MHD component. Also, the PCS often appears  in two different versions depending on whether the definition of the pressure tensor involves the absolute \cite{Cheng,FuPark} or relative \cite{KimEtAl,TaBrKi} velocity.

All of the nonlinear PCS models commonly found in the plasma physics literature suffer from the defect that they do not exactly  conserve  energy.  Indeed, exact energy conservation is lost when the assumption of a rarefied hot component is inserted as an approximation  in the equations of motion of the model.  Consequently, these models are not Hamiltonian field theories, ones that  are expected to have  noncanonical Poisson brackets akin to those introduced into  plasma physics in  \cite{MorrisonGreene} for MHD and \cite{Morrison2bis,Morrison3} for the Vlasov equation.  Since such a Hamiltonian structure occurs for all good plasma models, in their non-dissipative limit,(see \cite{Morrison3,morrison98,Morrison2005}), this would suggest there should be a Hamiltonian model for the PCS, and indeed this was shown to be the case in the recent literature \cite{HoTr2011,Tronci,MoTaTr2014}. This new model not only conserves energy,  but also conserves the cross-helicity invariants (which are also lost in the non-Hamiltonian case).

A main goal of the present paper is to compare the Hamiltonian and non-Hamiltonian PCS models, with emphasis on  linear stability analyses.  It is important to make clear that we are not arguing that dissipation is unimportant and that the Hamiltonian description is the most apt description of hybrid plasmas;  clearly this is not always the case -- collisional effects, albeit small, can give rise to important consequences, as is evident, { for instance,} from the massive body of reconnection studies in the literature.   Rather,  the goal here is to  investigate some consequences of nonphysical dissipation (or drive) that exists  in hybrid models when all the clearly identifiable physical dissipative terms are set to zero.  This fake dissipation may also be small, as is often the case for physical dissipation, but could lead to substantial yet erroneous consequences.  Indeed, we discover a spurious instability in the non-Hamiltonian model.

The remainder of the paper is organized as follows.  In Sec.~\ref{coupsch} we review the two hybrid coupling schemes:   the  CCS and PCS models are derived from first principles and general comments about their structure are made. This is followed in Sec.~\ref{linear} by a general treatment of the  linear problem for the incompressible PCS models expanded about  homogeneous equilibria in a uniform external magnetic field,  by integration over orbits.  This is followed, in Sec.~\ref{long}, by a study of the dispersion relation for  transverse disturbances parallel to the magnetic field.  It is in this special case that we compare the Hamiltonian and non-Hamiltonian PCS models and discover the spurious instability.  For completeness we also compare with the CCS models.  The dispersion relation is analyzed numerically and analytically and shown to have  a crossover to instability at high frequencies.   Next, in Sec.~\ref{trans}, comments are made about the behavior of perpendicular disturbances.  Finally, in Sec.~\ref{conclu} we summarize and conclude.  The paper contains two appendices that are included for completeness.  In Appendix \ref{app1} the noncanonical Poisson brackets for the Hamiltonian hybrid models are given, while  Appendix \ref{sec:disprel} records some details of our calculations leading to   the  dispersion relation used in Sec.~\ref{long}.

\section{Hybrid coupling schemes}
\label{coupsch}

Turning now to the two coupling schemes, we first consider the CCS, then the PCS.   

\subsection{Current-coupling schemes}

In order to derive the hybrid CCS model \cite{ParkEtAl}, one starts with the equations of motion for a multifluid plasma
in the presence of an energetic component. {Upon formally neglecting the vacuum permittivity (see e.g. \cite{Freidberg}), one writes}
\begin{align}
&\rho_s\left(\frac{\partial\bu_s}{\partial t}+  \bu_s\cdot\nabla \bu_s\right)
= -\nabla\mathsf{p}_s
\nonumber\\
&\hspace{3 cm}
+ \rho_sa_s \left(\bE+\bu_s\times\bB\right)\,,
\label{multi-fluid-momentum}\\
&
\frac{\partial\rho_s}{\partial t}+\nabla\cdot\left(\rho_s\bu_s\right)=0\,,
\label{multi-fluid-Vlasov}\\
&
\frac{\partial f}{\partial t}+\frac{\bp}{m_h}\cdot\frac{\partial f}{\partial \bx} 
+q_h\left(\bE+\frac{\bp}{m_h}\times\bB\right)\cdot\frac{\partial f}{\partial \bp}=0\,,
\nonumber\\
&
\nabla\times\bB=\mu_0 \bJ
\nonumber\\
&\hspace{1cm}= \mu_0\sum_s a_s\rho_s\bu_s+\mu_0\,a_h\bK\,,
\label{ampere}
\\
&
\frac{\partial\bB}{\partial t}=-\nabla\times\bE\,,
\label{faraday}
\\
&
0=\sum_s a_s\rho_s+q_hn
\,,\qquad
\nabla\cdot\bB=0
\label{neutrality}
\,,
\end{align}
where $a_s=q_s/m_s$ is the charge-to-mass ratio for the fluid species
$s$, while $\rho_s$ and $\bu_s$ are its mass density and velocity, respectively. {The symbol  $\mathsf{p}_s$, on the other hand, indicates the partial pressure of the fluid species $s$, which is assumed to be a function of $\rho_s$, through the relation $\mathsf{p}_s=\rho_s^2 \partial \mathcal{U}_s / {\partial} \rho_s$, with $\mathcal{U}_s (\rho_s)$ indicating  the corresponding specific internal energy.}

For simplicity, we consider from now on, the case in which the bulk plasma is composed by two species, one consisting of ions and the second one of electrons. It is customary to reduce the two-fluid system by neglecting  the inertia of the electron species (taking the limit ${m_2\to 0}$), thereby obtaining a one-fluid momentum equation.  With this assumption, summation of  Eqs.~\eqref{multi-fluid-momentum} for $s=1,2$ produces  
\begin{multline}
\label{TotMom}
\rho_1\bigg(\frac{\partial\bu_1}{\partial t}+\bu_1\cdot\nabla\bu_1\bigg)
=
\left(a_1\rho_1+a_2\rho_2\right)\bE 
\\
+\left(a_1\rho_1\bu_1+a_2\rho_2\bu_2\right)\times\bB-\nabla\mathsf{p}\,,
\end{multline}
{where $\mathsf{p}=\mathsf{p}_1+\mathsf{p}_2$}. 
Then, upon using Amp\`ere's law \eqref{ampere} and the quasineutrality relation of  \eqref{neutrality}, Eq.~(\ref{TotMom})  becomes
\begin{multline}
\label{veleq1}
\rho_1\bigg(\frac{\partial\bu_1}{\partial t}+\bu_1\cdot\nabla \bu_1\bigg)
=
- q_h  n \bE  
\\
+ \left(\bJ- \,a_h\bK\right)\times\bB-\nabla\mathsf{p}
\,,
\end{multline}
while Eq.~\eqref{multi-fluid-momentum} for the second species  yields   
\ben
\bE&=&-\bu_2\times\bB+
\frac1{a_2\rho_2}\nabla\mathsf{p}_2
\nonumber\\
&=& 
\frac1{a_2\rho_2}\left(a_1\rho_1\bu_1+a_h\bK- \bJ\right)\!\times\bB
+\frac1{a_2\rho_2}\nabla\mathsf{p}_2
\,.
\nonumber
\een
{Next, one imitates the derivation of ideal MHD  \cite{Freidberg} and assumes that $\bJ\times\bB$ and $\nabla\mathsf{p}_2$ are both negligible compared to the  Lorentz force $a_1\rho_1\bu_1\times\bB$. This step leads to} an Ohm's law of the form 
\begin{equation}
\bE
=
-\left(\frac{a_1\rho_1\bu_1+q_h n\bV}{a_1\rho_1+q_h n}\right)\!\times\bB\,, 
\label{Ohmslaw}
\end{equation}
where $\bV=m_h^{-1}\bK/n$ is the hot mean velocity. Note, this means magnetic flux is frozen-in at a velocity given by 
\begin{equation}
\bW=\frac{a_1\rho_1\bu_1+q_h n\bV}{a_1\rho_1+q_h n}\,.
\end{equation}
However,  if $\bV$and $\bu_1$ are comparable and $a_1\rho_1 \gg q_h n_h$, then one can replace \eqref{Ohmslaw} by the Ohm's law of  ideal MHD, 
\begin{equation}
\bE=-\bu_1\times\bB\,, 
\label{Ohmsideal}
\end{equation}
and the magnetic flux is then frozen into the MHD bulk flow. Finally, inserting \eqref{Ohmsideal} into Eqs.~\eqref{veleq1}, \eqref{multi-fluid-Vlasov}, and \eqref{faraday} yields the \underline{Hamiltonian CCS}: 
\begin{align}\label{cc-hybrid-momentum}
&\rho\left(\frac{\partial\bu}{\partial t}+\bu\cdot\nabla\bu\right)
= -\nabla\mathsf{p}
\\
&
\hspace{3 cm} + \left(q_h n\bu-\,a_h\bK + \bJ\right)\times\bB\,, 
\nonumber\\
& \frac{\partial\rho}{\partial t} +\nabla\cdot\left(\rho\bu\right)=0\,,  
\label{cc-hybrid-mass}
\\
&
\frac{\partial f}{\partial t}+\frac{\bp}{m_h}\cdot\frac{\partial f}{\partial \bx} 
+q_h\left(\frac{\bp}{m_h}-\bu\right)\times\bB\cdot\frac{\partial f}{\partial \bp}=0\,, 
\label{kinetic-CCS}
\\
&
\frac{\partial\bB}{\partial t}=\nabla\times\left(\bu\times\bB\right)
\label{cc-hybrid-end}
\,,
\end{align}
where the subscript 1 has been dropped. The system \eqref{cc-hybrid-momentum}-\eqref{cc-hybrid-end} is identical to the current-coupling hybrid scheme presented
in \cite{BeDeCh,ChWhFuNa,ParkEtAl}, except for the fact that the hot particle dynamics is governed by the Vlasov equation rather than gyrokinetic or drift-kinetic counterparts.

Note,  we always assume that the mean velocity $\bV=m_h^{-1}\bK/n$ of the energetic component is either very low or at {most}  comparable with the MHD fluid velocity $\bu$.  This is  consistent with the hypothesis of energetic particles,   since  the latter hypothesis involves the temperature rather than the mean velocity. Denoting the temperatures of the hot and fluid components by $T_h$ and $T_f$, respectively, we have   $T_h\gg T_f$ (see \cite{Cheng}).  With the  definition of the temperature  $T_h=\left(m_h/3nk_B\right){\int|\bv-\bV|^{2} f\,\de^3\bv}$ (where $k_B$ denotes Boltzmann's constant), the assumption on  the  energetic component amounts to an assumption on the trace of the second-order moment of the Vlasov density with no assumption on the mean velocity, which is actually low for hot particles close to isotropic equilibria \cite{TaBrKi}.


Notice that equation \eqref{cc-hybrid-momentum} involves the Lorentz force term $q_hn\bu\times\bB$, which should normally be negligible for consistency with the approximation $m_hn\ll\rho$, {which yields Eq.~\eqref{Ohmsideal} from Eq.~\eqref{Ohmslaw}}. A variant of the above CCS also exists \cite{ToSaWaWaHo}, which, {by virtue of the  approximation $m_hn\ll\rho$ neglects the term $q_hn\bu\times\bB$,  but on the other hand retains the term   $\bK\times\bB$ even though the two terms are in principle of the same order (as long as $\bV$ is comparable with $\bu$)} .

One can check directly that  the hybrid CCS model of  \eqref{cc-hybrid-momentum}-\eqref{cc-hybrid-end}  exactly  conserves  the following total energy:
\begin{multline}
\mathcal{E}=\frac1{2}\int\rho|\bu|^2\de^3\bx+\frac1{2m_h}\int f \left|\bp\right|^2\dvol 
\\
 +\int\! \rho\,\mathcal{U}(\rho)\,\de^3\bx
+\frac1{2\mu_0}\int \left|\bB\right|^2\de^3\bx
\,,
\label{Ham-preMHD}
\end{multline}
(see \cite{ToSaWaWaHo}) where $\mathcal{U}(\rho)$ is the  internal energy per unit mass, from which the pressure is determined by $\mathsf{p}=\rho^2\partial\mathcal{U}/\partial \rho$.   Moreover, this system  is  Hamiltonian,  with a noncanonical Poisson bracket  \cite{Tronci}   (recorded for completeness in Appendix \ref{app1}) and  it conserves the usual cross-helicity invariant $\int\!\bu\cdot\bB\,\de^3\bx$ \cite{HoTr2011}.

\subsection{Pressure-coupling schemes}

 Let us consider now the two models that use the PCS -- first the non-Hamiltonian version then the Hamiltonian {one}. 

\subsubsection{Non-Hamiltonian PCS}

Once the CCS has been obtained, the pressure-coupling scheme can be derived by computing the evolution of the total momentum
\begin{equation}\label{totmom}
\bM:=\rho\bu+\bK=:\rho \bU
\,,
\end{equation}
which gives (cf.\ Eq.~(1) of \cite{ParkEtAl})
\begin{equation}
\frac{\partial\bK}{\partial t}+\rho\left(\frac{\partial\bu}{\partial t}+\bu\cdot\nabla\bu\right)=-\nabla\cdot\Bbb{P}-\nabla\mathsf{p}+ \bJ\times \bB\,.
\end{equation}
Since $\partial_t\bK=\rho\,\partial_t(\bK/\rho)+(\operatorname{div}\bu)\bK/\rho$, inserting the assumption $\bK/\rho\ll\bu$ yields 
\begin{equation}
\rho\left(\frac{\partial\bU}{\partial t}+\bu\cdot\nabla\bu\right)=-\nabla\cdot\Bbb{P}-\nabla\mathsf{p} 
+\bJ\times \bB\,. 
\end{equation}
  Then, upon writing $\bU\sim\bu$, we obtain the system with the \underline{non-Hamiltonian PCS}:  
\begin{align}
&\rho\left(\frac{\partial\bU}{\partial t} +\bU\cdot\nabla\bU\right)=-\nabla{\sf p}-\nabla\cdot\Bbb{P}
+ \bJ \times \bB
\label{PCS-FuPark}
\\
& \frac{\partial\rho}{\partial t} + \nabla\cdot\left(\rho\bU\right)=0 
\label{pc-hybrid-mass}
\\
&
\frac{\partial f}{\partial t}+\frac{\bp}{m_h}\cdot\frac{\partial f}{\partial \bx}+q_h\left(\frac{\bp}{m_h}-\bU\right)\times\bB\cdot\frac{\partial f}{\partial \bp}=0
\label{kinetic-PCS}
\\
&
\frac{\partial\bB}{\partial t}=\nabla\times\left(\bU\times\bB\right)
\label{pc-hybrid-end}
\,,
\end{align}
Notice that Eq.~\eqref{PCS-FuPark} is identical to the bulk momentum equation of  the hybrid PCS of Fu and Park (see equations (1) in \cite{FuPark,FuParketAl}), which also includes \eqref{pc-hybrid-mass} and \eqref{pc-hybrid-end} (while replacing Vlasov dynamics by its gyrokinetic approximation). 
 Analogous PCS models with the same fluid equation \eqref{PCS-FuPark} have been formulated by Cheng \cite{Cheng} (see equation (1) therein) and Park \emph{et al.}\  \cite{ParkEtAl} (see equation (3) therein). In some situations, the tensor \eqref{presstensor} in \eqref{PCS-FuPark} is replaced by the relative pressure tensor $\widetilde{\Bbb{P}}=m_h^{-1}\int\left(\bp-m_h\boldsymbol{V}\right)\left(\bp-m_h\boldsymbol{V}\right)f\,\de^3\bp$ (e.g.,  the PCS model proposed by Kim, Sovinec and Parker \cite{Kim,KimEtAl,TaBrKi}).

All the above mentioned PCS models suffer from not conserving the total energy exactly. Indeed, if we assume that the total energy is still given by  \eqref{Ham-preMHD}, Eqs.~\eqref{PCS-FuPark}-\eqref{pc-hybrid-end} give $\dot{\mathcal{E}}=\int\bU\cdot\partial_t\bK\,\de^3\bx$, so that the total energy would only be nearly conserved if $\partial_t\bK$ is small. Under this assumption, the CCS and the PCS are completely equivalent, since \eqref{kinetic-CCS} yields
\begin{equation}
\frac{\partial\bK}{\partial t}=-\nabla\cdot\Bbb{P}+\,a_h\bK\times\bB-
q_h n\bu\times\bB
\,.
\end{equation}
However, the assumption that $\partial_t\bK$ is negligible is \emph{not} compatible with \eqref{kinetic-CCS}, since the time variation of $\bK$ may indeed play a role in the general case.

\subsubsection{Hamiltonian PCS} 

 The issue of exact energy conservation was raised in \cite{Tronci}, where an alternative Hamiltonian version of the PCS was presented. Besides conserving the energy \eqref{Ham-preMHD} exactly, this model possesses a Poisson bracket structure, which was derived by using well established Hamiltonian techniques in geometric plasma dynamics \cite{morrison98,MaWe1,MaWeRaScSp,Morrison2005,Morrison2,Morrison3,Morrison2bis,MorrisonGreene,Spencer,SpKa}.

In order to derive the Hamiltonian PCS model of  \cite{Tronci}, one expresses  the Hamiltonian structure of the CCS \eqref{cc-hybrid-momentum}-\eqref{cc-hybrid-end} in terms of the total momentum $\bM$ in \eqref{totmom}. Then, instead of replacing  $\bU\sim\bu$ (arising from the original assumption  $m_hn\ll\rho$) in the equations of motion, one replaces $\bU\sim\bu$ directly in \eqref{Ham-preMHD} and derives the equations of motion from the Poisson bracket structure written in terms of $\bM$ (see Appendix \ref{app1}). This procedure ensures that the chosen energy functional is always preserved, as long as no approximations are made on the Poisson bracket. At this point, one obtains the following set of equations for the \underline{Hamiltonian PCS}: 
\begin{align}
&\rho\left(\frac{\partial\bU}{\partial t}+\bU\cdot\nabla\bU\right)=-\nabla{\sf p}-\nabla\cdot\Bbb{P}
+\bJ\times\bB
\label{PCS-FuPark2}
\\
\label{hybridMHD2}
&\frac{\partial f}{\partial
t}+\left(\frac{\bp}{m_h}+\boldsymbol{U}\right)\cdot\frac{\partial
f}{\partial \bx} 
\\
&\hspace{.5cm} +\Big[{\bp}\times\big(a_h \bB-\nabla\times\bU\big) 
-\bp\cdot\nabla\bU\Big]\cdot\frac{\partial f}{\partial\bp}=0\,,
\nonumber\\
\label{hybridMHD3}
&\frac{\partial \rho}{\partial t}+ \nabla\cdot (\rho\,\bU)=0
\,,\qquad\quad\!
\frac{\partial \bB}{\partial t}=\nabla\times\left(\bU\times\bB\right)
\,.
\end{align}
We see that the fluid equation \eqref{PCS-FuPark2} is identical to the corresponding equation \eqref{PCS-FuPark} of the non-Hamiltonian model.
However, in the Hamiltonian model, hot particles move with the relative velocity $\bU+\bp/m_h$ so that the term $\nabla\bU\cdot\bp=\bp\cdot\nabla\bU+ \bp\times(\nabla\times\bU)$ appears as  an inertial force. Consequently, both Hamiltonian and non-Hamiltonian PCS'  possess the same static equilibria, although the dynamics in the vicinity of these  equilibria may be very different, depending on the particular situation  under consideration. Also, we notice that, unlike the CCS of  \eqref{cc-hybrid-momentum}-\eqref{cc-hybrid-end} and the non-Hamiltonian PCS of  \eqref{PCS-FuPark}-\eqref{pc-hybrid-end}, the Hamiltonian equations  of \eqref{PCS-FuPark2}-\eqref{hybridMHD3} involve a nontrivial kinetic-fluid coupling,  even in {the} absence of magnetic fields.

\section{Linearized incompressible PCS}
\label{linear}

In this section, we consider the linearized equations of motion for the incompressible limit  (\emph{e.g.} $\nabla\cdot\bU=0$)  of both Hamiltonian and non-Hamiltonian PCS'. For the sake of simplicity, we shall set all physical constants to unity (including $m_h$, so that $\bp=\bv$), although we shall restore them at a later time. 

\subsection{Equations of motion}

 Upon following the standard procedure, we linearize each variable as ${A}={A}_0+{A}_1$, so that the subscripts ``0'' and ``1'' denote an equilibrium and its perturbation, respectively. As a result, we obtain 
\begin{align}\label{hybridMHD1--lin}
&\frac{\partial \bU_1}{\partial t}=-\nabla{\sf p}_1-\nabla\cdot \int\! \bv\bv f_1\,\de^3\bv
+ (\nabla\times\bB_1)\times \bB_0\,, 
\\\label{hybridMHD2--lin}
&\frac{\partial f_1}{\partial
t}+\bv\cdot\frac{\partial
f_1}{\partial \bx}+\bv\times\bB_0\cdot\frac{\partial
f_1}{\partial \bv}= 
\\
&\hspace{2.7  cm} f_0'\left(\alpha
\bv\bv: \nabla\boldsymbol{U}_1+\beta\bv\cdot\bU_1\times \bB_0\right)\,,
\nonumber
\\\label{hybridMHD3--lin}
&
\frac{\partial \bB_1}{\partial t}=\nabla\times\left(\bU_1\times\bB_0\right)
\,,\qquad
\nabla\cdot\bU_1=0
\,,
\end{align}
where the parameters $\alpha$ and $\beta=1-\alpha$ are inserted so that $\alpha=1$ gives the linearized Hamiltonian model, while $\alpha=0$ gives the linearized non-Hamiltonian model. In this way, it is clear that the $\alpha$-terms identify the Hamiltonian model, while the $\beta$-terms identify its non-Hamiltonian counterpart. {In Eqs.~(\ref{hybridMHD1--lin})-(\ref{hybridMHD3--lin}) we assumed} a static equilibrium so that $\bU_0\equiv 0$ and $\mathsf{p}_0\equiv 0$. Also, we consider a uniform magnetic field $\bB_0$ (aligned with the $z$-axis) and an isotropic equilibrium for the energetic component, so that $f_0=f_0(v^2/2)$ with $v^2=|\bv|^2$. Notice that the special case $\bB_0=0$ yields free transport for the hot particles, in the case of the non-Hamiltonian model ($\alpha=0$). Conversely, the Hamiltonian model ($\alpha=1$) retains the fluid velocity terms in the kinetic equation, even in the absence of the magnetic field ($\bB_0=0$). 

Notice that, although here we have chosen a Vlasov equilibrium of the form $f_0=f_0(v^2/2)$, other more realistic choices are also available. For example, in actual hybrid simulations of the non-Hamiltonian PCS in fusion devices,  toroidal symmetry is involved and the use of special equilibrium profiles becomes necessary \cite{TaBrKi}. Another example arises in reversed field pinch plasmas, in which finite Larmor radius effects allow for equilibria of the type $f_0=f_0(\bx,v^2)-\omega_c^{-1}\nabla f_0\cdot\bv\times\bB$ (where $\omega_c= q_hB_0/m_h$) (see \cite{Kim}).  On the other hand, the aim  here is not to enter  into detailed features of particular fusion devices, but  to provide  insight  into model differences.    Therefore, we focus on  distributions of the type $f_0=f_0(v^2/2)$.

Assuming perturbations  {varying} as  ${A}_1=\widetilde{{A}}_1 e^{i(\mathbf{k\cdot x}-\omega t)}$ gives 
\begin{align}\label{hybridMHD1--Fo}
&\omega\tU=\bk\,\tp+\int\! (\bk\cdot\bv)\bv \tf\,\de^3\bv
+\bB_0\times(\bk\times \tB)\,,
\\\label{hybridMHD3--Fo}
&
-\omega\tB=(\bk\cdot\bB_0)\tU
\,,\qquad
\bk\cdot\tU=0
\,. 
\end{align}
 Dotting Eq.~(\ref{hybridMHD1--Fo}) by $\bk$ and  using Eqs.~(\ref{hybridMHD3--Fo}) yields
\[
\tp=-\frac1{\ |\bk|^2}\int\!(\bk\cdot\bv)^{2\,}\tf\,\de\bv+\frac1\omega\,(\bk\cdot\bB_0)(\tU\cdot\bB_0)\,, 
\]
and the velocity equation becomes
\begin{multline}
\omega\tU=-\int\!\left[\frac{(\bk\cdot\bv)^2}{\ |\bk|^2}\bk-(\bk\cdot\bv)\bv\right]\tf\,\de\bv
\\
+\frac1\omega\,(\bk\cdot\bB_0)^2\,\tU
\end{multline}
or, upon rearranging the various terms,
\begin{align}
\label{velocityrelation}
\tU
=& \frac{\omega}{(\omega^2-\bk\cdot\bB_0)^2}\,\bigg[\mathbf{1}- \frac{\bk\bk}{\,{|\bk|}^2}\bigg](\bk\cdot{\widetilde{\Bbb{P}}}_1)
\,,
\end{align}
where $\widetilde{\Bbb{P}}_1=\int\!\bv\bv\,\tf\,\de^3\bv$ and $
\mathbf{1}- \bk\bk/|\bk|^2$ projects transverse to $\bk$.


It remains to express $\tf$ in terms of $\tU$  in order to obtain the dispersion relation. This step is performed in Sec.~\ref{sec:Vlasovsol}, but first the next section contains some relevant properties of the linearized equation of Vlasov kinetic moments.

\subsection{Remarks on linearized moment dynamics\label{Sec:moments}}

Before analyzing the linear dynamics, it is of interest to explore some  roles played by the $\alpha$ and  $\beta$ terms. To this end it is useful to introduce equilibrium moments
\[
\left(A_n^{(0)}\right)_{i_1 i_2\dots i_n}=\int\! v_{i_1}\dots v_{i_n}\,f_0\,\de^3\bv
\,.
\]
Since $f_0=f_0(v^2/2)$,  we notice that $A_{2n+1}^{(0)}=0$.
Then, Eq.~\eqref{hybridMHD2--lin} leads to the following conclusions about the equations of motion for the kinetic moments ${A}_k^{(1)}=\int v^kf_1\,\de^3\bv$:
\begin{itemize}
\item the $\alpha$-term (of the Hamiltonian model) contributes only to moments of even order $2n+2$ (\emph{e.g.} the pressure tensor $\Bbb{P}_1$), i.e.,  the $\alpha$-term does not contribute  to the dynamics of odd-order moments;
\item the $\beta$-term (of the non-Hamiltonian model) contributes only to moments of odd order $2n+1$ (\emph{e.g.} the averaged momentum $\bK_1$), i.e.,  the 
$\beta$-term  does not contribute  to the dynamics of  even-order moments;
\item  the first three moments obey the equations
\begin{align*}
&\partial_t n_1+\nabla\cdot\bK_1=0
\\
&\partial_t\bK_1+\nabla\cdot\Bbb{P}_1-\bK_1\times\bB_0=-\beta n_0\,\bU_1\times\bB_0
\\
&\partial_t\Bbb{P}_1+\nabla\cdot A_3^{(1)}+\big[\widehat{B}_0,\Bbb{P}_1\big]=-2\alpha\big((\Bbb{P}_0\cdot\nabla)\bU_1 
\\
&\hspace{3.5cm} +((\Bbb{P}_0\cdot\nabla)\bU_1)^T\big)
\,,
\end{align*}
where $[\cdot,\cdot]$ denotes matrix commutator and we defined  the hat operator  by $\widehat{w}\mathbf{a}:={\boldsymbol{w}}\times\mathbf{a}$ (for any two vectors $\boldsymbol{w}$ and $\mathbf{a}$, so that $\widehat{w}_{ih}=-\epsilon_{ihk}{w}_k$ is  an antisymmetric matrix).
\end{itemize}
Thus,  there is no contribution of the $\alpha$-term to the linearized fluid moments (i.e. zeroth and first order moments) of the kinetic component: indeed, the $\alpha$-term disappears  in the linearized dynamics of the fluid closure for the hot particles. The $\alpha$-term contributes only to the dynamics of perturbed moments at  even order ({\it e.g.} the pressure tensor).
On the other hand, the $\beta$-term contributes only to the dynamics of perturbed moments with odd order, e.g. it has a non-zero contribution to the first-order moment (which plays a crucial role in the current-coupling scheme).

\subsection{Solution of the linearized Vlasov equation\label{sec:Vlasovsol}}

Now  we solve the linearized kinetic equation \eqref{hybridMHD2--lin} in terms of the fluid velocity $\bU_1$.  This is done by  invoking  the method of characteristics (integrating over orbits) as is standard in  plasma physics texts (e.g.\  \cite{KrallTravelpiece})  for the Maxwell-Vlasov system. Following this standard method  yields the solution of the Vlasov equation in the form
\begin{multline}
f_1=\int_{-\infty}^t\!f_0'\left(\alpha\nabla\bU^*_1:\bv^*\bv^*-\beta\bU^*_1\cdot\bv^*\times\bB_0\right)\de t^*
\nonumber\\
 +f_1(\bx^*(-\infty),\bv^*(-\infty),-\infty)
\,,
\nonumber
\end{multline}
where  $f_0'$ means derivative of $f_0$ with respect to its  argument 
${v^*}^2/2$ and the variables $(\bx^*(t^*),\bv^*(t^*),t^*)$ satisfy 
\[
\dot{\bx}^*(t^*)=\bv^*(t^*)\,,\qquad\dot{\bv}^*(t^*)=\bv^*(t^*)\times\bB_0\,,
\]
with $\bx^*(t)=\bx$, $\bv^*(t)=\bv$ {and the dot indicating the derivative with respect to the evolution parameter $t^*$.} 

Then, upon introducing the notation $\omega_{_{\!B_0}}=|\bB_0|$ (i.e. the cyclotron frequency, upon restoring physical quantities) and the planar rotation
\[
\mathcal{R}(\tau)=\exp(\tau\widehat{B}_0)
=\left(\begin{array}{ccc}
\cos(\omega_{_{\!B_0}}\!\tau)&  -\sin(\omega_{_{\!B_0}}\!\tau) & 0\\
\sin(\omega_{_{\!B_0}}\!\tau) &\cos(\omega_{_{\!B_0}}\!\tau) & 0\\
0 & 0 & 1
\end{array}\right)
\]
and 
its antiderivative ${A\mathcal{R}}(\tau)$, we have
\[
\bx^*={A\mathcal{R}}(\tau)\bv+\omega_{{\!B_0}}^{-2\,}\bB_0\times\bv+\bx\,, 
\]
together with 
$\bv^*=\mathcal{R}(\tau)\bv$  and $\tau=t^*-t$.
Upon Fourier-transforming in the spatial variable,  we obtain 
\begin{align*}
\tf=&\int_{-\infty}^0\!f_0'\left(i\alpha(\bk\cdot\bv^*)(\tU\cdot\bv^*)+\beta\tU\cdot\bB_0\times\bv^*\right)
\nonumber\\
&\hspace{4 cm} \times e^{i\left(\bk\cdot\bX-\omega\tau\right)}\de\tau
\\
=&\int_{-\infty}^0\!f_0'\left(i\alpha(\bk\cdot\mathcal{R}(\tau)\bv)\tU-\beta\bB_0\times\tU\right)\cdot\mathcal{R}(\tau)\bv
\nonumber\\
&\hspace{4 cm}\times  e^{i\left(\bk\cdot\bX-\omega\tau\right)}\de\tau\,, 
\end{align*}
where  $\bX:=\mathbf{x}^*-\mathbf{x}$ and recall
$v^*_z=v_z$ and $z^*(\tau)=v_z\tau + z$,
so that $\mathcal{R}\bB_0=\bB_0$ and $(\mathcal{R}\tU)_z=\widetilde{U}_{1 z}$. Thus, since $\mathcal{R}$ is a rotation,  
\begin{multline}
\label{Vlasovsolution}
\tf
=\int_{-\infty}^0\frac{\partial f_0}{\partial \bv}\cdot
\Big(i\alpha(\bk\cdot\mathcal{R}(\tau)\bv)\mathcal{R}^T(\tau)\tU
\\
-\beta\bB_0\times\mathcal{R}^T(\tau)\tU\Big)\, 
 e^{i\left(\bk\cdot\bX-\omega\tau\right)}\,\de\tau\,.
\end{multline}
Finally,  upon recalling the definition  $\widehat{w}\mathbf{a}:={\boldsymbol{w}}\times\mathbf{a}$ of hat operator, the velocity equation becomes
\begin{multline}
\left(\omega^2-(\bk\cdot\bB_0)^2\right)\tU=
 \label{Urelation} \\
\omega\bigg[\mathbf{1}- \frac{\bk\bk}{\,{|\bk|}^2}\bigg]\! \iint_{-\infty}^0\!(\bk\cdot\bv)\!
\bigg[
\mathcal{R}\!
\Big(i\alpha(\bk\cdot\mathcal{R}\bv)\frac{\partial f_0}{\partial \bv} 
\\ 
+\beta\widehat{B}_0 \frac{\partial f_0}{\partial \bv}\Big)\!\cdot\tU
\bigg]
\!\bv e^{i\left(\bk\cdot\bX-\omega\tau\right)}\,\de\tau\,\de\bv\,.
\end{multline}
and the dispersion relation is
\begin{multline}
\label{disprelgen}
\operatorname{det}\!\bigg(k^2\!\left((\bk\cdot\bB_0)^2-\omega^2\right)\boldsymbol{1}
\\
-\omega \widehat{k}^2\!\iint_{-\infty}^0\!(\bk\cdot\bv)\,\bv\!\left(i\alpha(\bk\cdot\mathcal{R}\bv)\boldsymbol{1}+\beta\widehat{B}_0\right)
\\
\times \, 
\mathcal{R}\frac{\partial f_0}{\partial \bv} e^{i\left(\bk\cdot\bX-\omega\tau\right)}\,\de\tau\,\de\bv
\bigg)=0
\end{multline}
At this point, one may write the {general} dispersion relation explicitly. However, we study the linearized system in {two particular} cases where $\bk_\perp=0$ and $k_z=0$, which we turn to in the next sections.

\section{Disturbances with $\bk_\perp=0$}
\label{long}

Now we specialize  the preceding results to  the special case $\bk_\perp=0$, thus giving the dispersion relation for  parallel propagating  transverse disturbances with  $\mathbf{k}\cdot \widetilde{\mathbf{E}}_1=-\mathbf{k}\cdot\tU\times\mathbf{B}_0=0$.

\subsection{Dispersion relation}

After setting  $\bk_\perp=0$ in Eq.~\eqref{Urelation}  it is useful to compute the perturbed moment quantities  $\tK$ and $\bk\cdot{\widetilde{\Bbb{P}}}_1$.  Notice that $\bk_\perp=0$ implies $k_z\widetilde{U}_{1 z}=0\Rightarrow\widetilde{U}_{1 z}=0$.  We compute $\tK$ by taking the first-order moment of \eqref{Vlasovsolution}.
Upon integrating by parts and recalling $\mathcal{R}^T\bk=\mathcal{R}\bk=\bk$, we have
\begin{align*}
\tK=&\int\bv \,\tf(\bv)\,\de\bv
\\
=&\int\!\bv  \frac{\partial f_0}{\partial \bv_\perp}\cdot
\int_{-\infty}^0\Big(i\alpha k_z v_z\mathcal{R}^T(\tau)\tU
 \\
&\hspace{1cm} - \beta\bB_0\times\mathcal{R}^T(\tau)\tU  \Big)  
e^{i\left(k_z v_z\tau-\omega\tau\right)}\,\de\tau \de\bv_\perp\de v_z
\\
=&-\int
\int_{-\infty}^0f_0\Big(i\alpha k_z v_z\mathcal{R}^T(\tau)\tU
\\
&\hspace{1cm}  -\beta\bB_0\times\mathcal{R}^T(\tau)\tU\Big)
 e^{i\left(k_z v_z\tau-\omega\tau\right)}\,\de\tau\de\bv
\\
=&-\int
\int_{-\infty}^0\bar{f_0}(v_z^2/2)\Big(i\alpha k_z v_z\mathcal{R}^T(\tau)\tU
 \\
&\hspace{1cm} 
-\beta\bB_0\times\mathcal{R}^T(\tau)\tU\Big) e^{i\left(k_z v_z\tau-\omega\tau\right)}
\,\de\tau\de v_z\,.
\end{align*}
Hence, 
\[
\widetilde{K}_{1 z}=0
\,.
\]
The above relation means that the momentum perturbation $\tK$ is coplanar with $\tU$, i.e.,  ${\tK\times\tU}=0$ and therefore the density perturbation vanishes, since $\widetilde{n}_1=-\bk\cdot\tK/\omega\equiv0$.
Here, we have introduced the notation $\bar{f_0}(v_z^2/2)=\int\! f_0\,\de\bv_\perp$. 
Notice that, by proceeding analogously, we have
\begin{multline*}
\bk\cdot{\widetilde{\Bbb{P}}}_1=-
\iint_{-\infty}^0\bar{f_0}(v_z^2/2)\,k_z v_z\Big(i\alpha k_z v_z\mathcal{R}^T\tU
\\
 -\beta\bB_0\times\mathcal{R}^T\tU\Big) 
e^{i\left(k_z v_z-\omega\right)\tau}\,\de\tau\de v_z
\end{multline*}
At this point one needs to compute the matrix integral
\[
\mathcal{A}=\int_{-\infty}^0 e^{i(k_z v_z-\omega)\tau}\mathcal{R}(\tau)\,\de \tau\,,
\] 
whose components are
\begin{align}\label{A11}
\mathcal{A}_{11}&=\mathcal{A}_{22}=
-i\frac{k_z v_z-\omega}{(k_z v_z-\omega)^2-\omega_{{\!B_0}}^2}
\,,
\\\label{A12}
\mathcal{A}_{12}& =-\mathcal{A}_{21}
=
\frac{\omega_c}{(k_z v_z-\omega)^2-\omega_{{\!B_0}}^2}
\,,
\\\nonumber
\mathcal{A}_{33}&=
-\frac{i}{k_z v_z-\omega}
\,.
\end{align}
 In  the above equations,  the integrations are defined for  $\operatorname{Im}(\omega)>0$;   following the standard  procedure  the solution is  extended to the lower complex plane by analytical continuation.
In conclusion, we have
\begin{align*}
\tK=&\
-\int\bar{f_0}(v_z^2/2)\left(i\alpha k_z v_z\boldsymbol{1}-\beta\widehat{B}_0\right)\mathcal{A}^T\tU \,\de v_z\,, 
\end{align*}
where we recall  $\widehat{B}_{0ij}=-\epsilon_{ijk}{B_0}_k$ and $(\mathcal{A}\tU)_z=0$, so that ${\widetilde{K}_{1 z}=0}$. 
Similarly, we obtain
\begin{equation}\label{pressureterm}
\bk\cdot{\widetilde{\Bbb{P}}}_1=
-\int k_z v_z \bar{f_0}\left(i\alpha k_z v_z\boldsymbol{1}-\beta\widehat{B}_0\right)\mathcal{A}^T\tU \,\de v_z
\,. 
\end{equation}

Once the moment quantities $\tK$ and $\bk\cdot{\widetilde{\Bbb{P}}}_1$ are written explicitly, one is ready to write the dispersion relation for  the case $\bk_\perp=0$.  Inserting the relation \eqref{pressureterm} into \eqref{velocityrelation} yields
\begin{multline}\label{pre-disprel}
({k_z^2b^2-\omega^2})\tU
= {\omega}\!\int\!k_z  v_z \bar{f_0}\Big(i\alpha k_z v_z \mathcal{A}^T\tU
\\ -\beta \bB_0\times\mathcal{A}^T\tU\Big) \de v_z\,.
\end{multline}
Notice, since $\widetilde{U}_{1 z}=({\mathcal{A}^T\tU})_z = ({\bB_0\times\mathcal{A}^T\tU})_z=0$, this relation only possesses planar components. 

At this point, direct algebraic computations on the above relation give the following dispersion relation:
\begin{multline}
 \label{disprel}
 D^{\pm}(\omega,k_z):= \omega^2-k_z^2 v_A^2
 +\omega(\alpha\omega\mp\omega_c) n_{0} \bigg(1 +
 \\
(\omega\mp\omega_c)\int_{-\infty}^{\infty}\!\frac{F}{k_z v_z-\omega\pm\omega_c}\,\de v_z\bigg)=0\,,
\end{multline}
where all physical constants have been restored:  $\mathbf{B}_0=B_0 {\z}$,  $\omega_c=q_hB_0 /m_h$ is the cyclotron frequency of the energetic component,    $v_A=B_0 /\sqrt{\mu_0 \rho}$  indicates the Alfv\'en speed based on the constant equilibrium magnetic field $B_0$ and the constant bulk mass density $\rho$, and  $F:= \bar{f}_0/n_0$ with  $n_0=\int_{-\infty}^{\infty} \bar{f_0} d v_z$, so that $n_0$  is a dimensionless number indicating the ratio between the equilibrium  mass  density of the energetic component and that of the bulk component. (For details of this calculation see Appendix \ref{sec:disprel}.)


\subsection{Analysis of   dispersion relation}
\label{sec:analysis}

Next we analyze the dispersion relation  \eqref{disprel}.  Recall,  if one sets $\alpha=1$ in (\ref{disprel}), one obtains the dispersion relation for the Hamiltonian model, whereas $\alpha=0$ gives that for the non-Hamiltonian model.
We will see that neglecting the terms that make the starting model Hamiltonian  leads  to important qualitative differences  in  the stability properties. 

Some consequences of the dispersion relation \eqref{disprel} are immediate. {In the absence} of energetic particles ($n_0=  0$), one recovers the dispersion relation $\omega=\pm v_A k_z$,  describing  Alfv\'en waves propagating along the $z$-direction.  Next, assume `cold hot' particles, i.e., the case where $F$ is the Dirac delta function $\delta(v_z)$.  In this case the hot particle contribution again vanishes, indicating that thermal effects are necessary to influence the Alfv\'en waves for both the Hamiltonian and non-Hamiltonian models.

\begin{figure}[htb]
\centering
\subfigure[{\footnotesize \  }]{\includegraphics[width=0.45\textwidth]{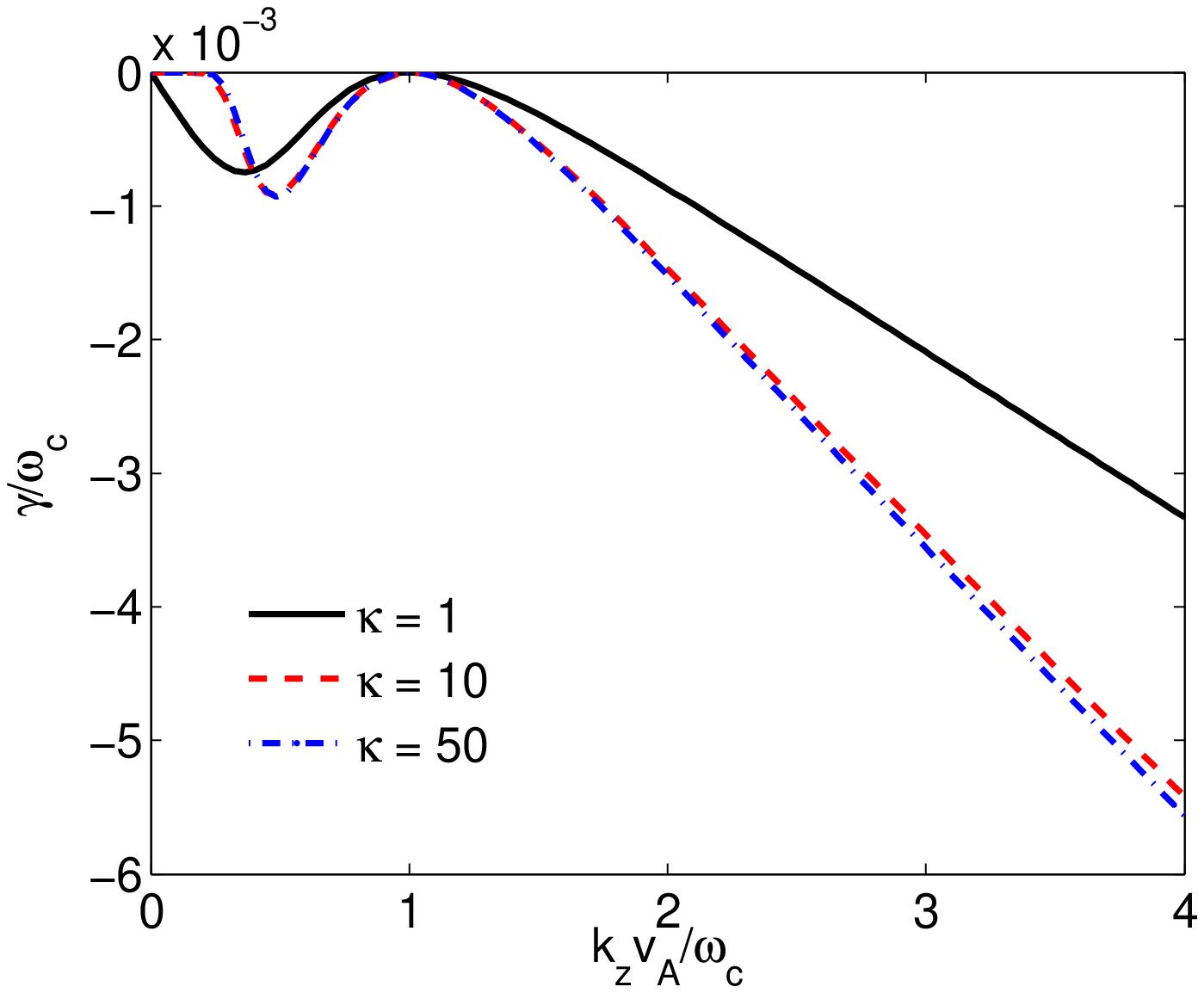}
\label{fig:gamkH}
}
\subfigure[{\footnotesize \  }]{\includegraphics[width=0.45\textwidth]{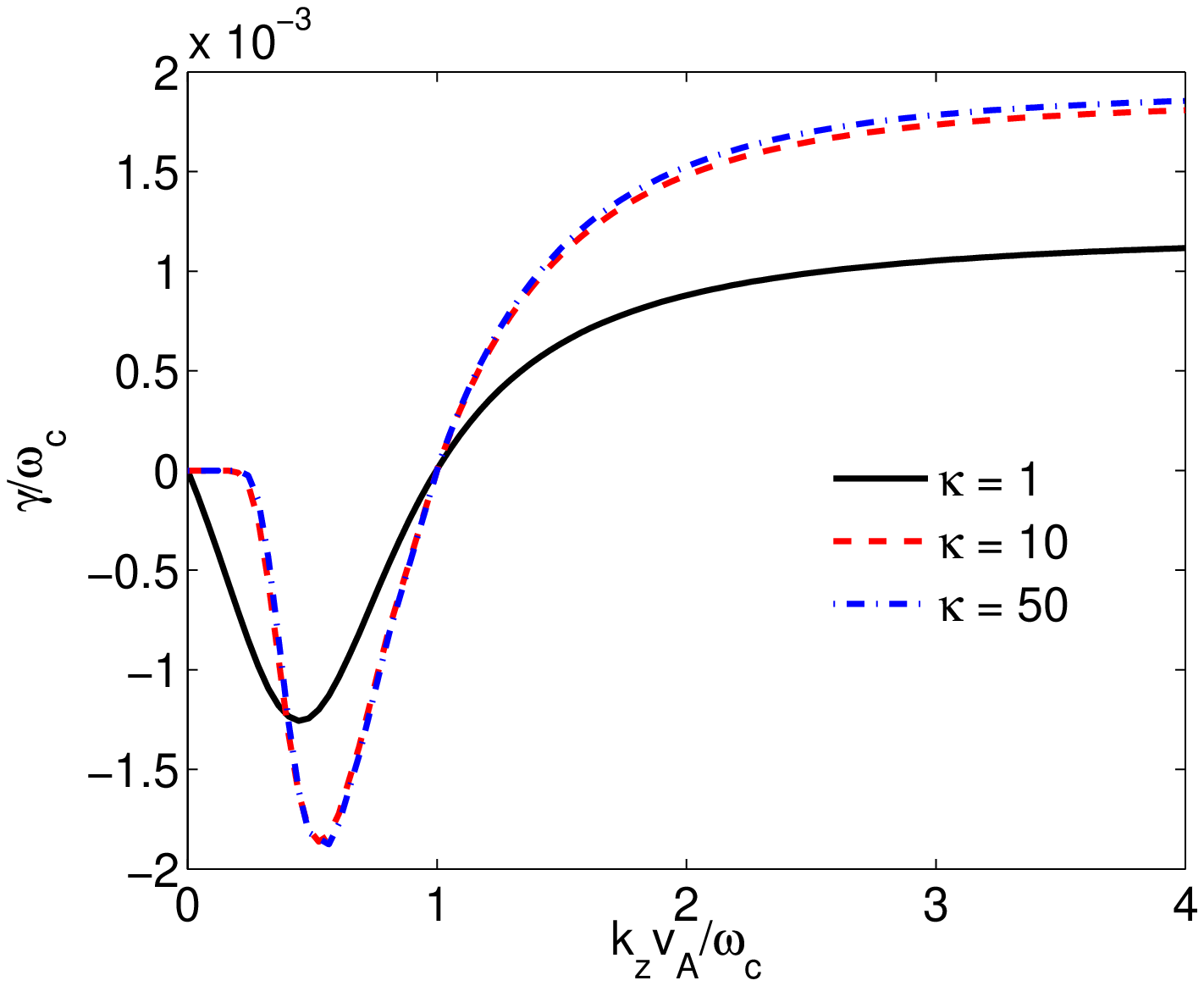}
\label{fig:gamkNH}
}
\caption[sm]
{Plots of the normalized damping/growth rates vs.\  wavenumber $k_z$ for the PCS using the kappa distribution of Eq.~\eqref{kappa} for different values of $\kappa$.  Here $n_0=5 \times 10^{-3}$ and {$v_0/v_A=1.2$}. Panel  \ref{fig:gamkH} corresponds to the Hamiltonian PCS, which shows the expected damping, while Panel   \ref{fig:gamkNH} corresponds to the non-Hamiltonian PCS, which depicts the spurious instability for frequencies above $\omega_c$. 
}
\label{fig:ga-k}
\end{figure}

To further analyze the two PCS' consider Fig.~\ref{fig:ga-k}, where  results are displayed from a numerical solution of the dispersion relation  of \eqref{disprel}  for the kappa distribution,
\begin{equation}
f^{(\kappa)}_0=\frac{n_0}{(\pi\kappa v_0^2)^{3/2}}
\frac{\Gamma(\kappa+1)}{\Gamma(\kappa-1/2)}
\left(1+\frac{v^2}{\kappa v_0^2}\right)^{-(\kappa+1)}\,.
\label{kappa}
\end{equation}
Here   $v_0$  reflects the thermal velocity $v_{\rm th}=\sqrt{k_B T/m_h}$.   Note, for large values of $\kappa$ the $\kappa$-distribution is indistinguishable from the Maxwellian (see, e.g., \cite{Podesta}), and we have verified this by direct calculation by comparing the two for $\kappa=50$. In Fig.~\ref{fig:ga-k},    the imaginary part of the frequency,  $\gamma$,  is plotted against the wavenumber, $k_z$, suitably normalized, for the counter polarization, i.e., for $D^{+}$, which gives the weakest damping.    In this figure we  consider a hydrogen bulk plasma with a particle density of $10^{14} \text{cm}^{-3}$, a magnetic field of 35 kgauss,  and {a alpha particle  component} with temperature of \makebox{3.6 MeV} and fractional density $n_0=5 \times 10^{-3}$.  This gives  {$v_{0}/v_A=1.2$} and  $\omega_c=8.4\times10^7 {\text{Hz}}$.  Since $n_0\ll1$, the real part of the frequency  corresponds nearly to the Alfven wave, i.e.,  $\omega_r\approx k_{z}v_A$, so it is not plotted.    Figure \ref{fig:gamkH} depicts $\gamma$  for the Hamiltonian PCS,   while Fig.~\ref{fig:gamkNH} shows the corresponding plot for the non-Hamiltonian PCS. The same behavior was found by varying $n_0$ within the range $n_0\approx 10^{-3}-10^{-1}$, in agreement with the relations \eqref{weakdamping}-\eqref{al1} below, obtained by the small growth rate expansion.

The first observation to make is that both the Hamiltonian and usual non-Hamiltonian models have similar behavior  for low frequencies.   This is to be expected,  since the non-Hamiltonian pressure coupling model was first developed to explore linear low frequency behavior.  In fact, for example in \cite{Cheng},  low frequency `$\delta W$' type arguments were given that indicate stability in this frequency regime, which is consistent with the figures. 

However, upon examinination of  Fig.~\ref{fig:ga-k} for larger values of $k_{z}$ or  $\omega_r \approx k_{z}v_A$,    we see that Figs.~\ref{fig:gamkH}   and  \ref{fig:gamkNH} differ as  $\omega_r$ approaches and exceeds $\omega_c$.     Most significantly, we see that the non-Hamiltonian PCS possess an instability for frequencies greater than $\omega_c$,  as is clearly evident  in Fig.~\ref{fig:gamkNH}.   Since the equilibrium we are considering has no available free energy, in either the bulk or  in the hot particles, this instability must be nonphysical and  reflects the  lack of energy conservation in  non-Hamiltonian PCS.  For the Hamiltonian PCS displayed in Fig.~\ref{fig:gamkH}, the system damps as expected.  
The hot particles provide Landau damping, in much the way one expects for electron Landau damping of Alfv\'en and whistler modes, with the mode at $\omega_c$ being undamped for one of the polarizations. 

For $\kappa=1$, it is easily shown by  residue calculus that 
\[
\int_{-\infty}^{+\infty}\!\frac{F}{k_z v_z-\omega\pm\omega_c}\,\de v_z
=
-\frac{1}{ik_z v_0+\omega\mp\omega_c}
\,,
\]
{
where we recall $F=\bar{f}_0(v_z^2)/n_0$ and $\bar{f}_0(v_z^2)$ is obtained from  Eq.~(44) for $\kappa=1$ (i.e. $f_0^{(1)}(v^2)$), upon integrating out the perpendicular components of the velocity.} 
  Thus    \eqref{disprel} becomes
\begin{equation*}
\omega^2-k_z^2 v_A^2
\\
+in_{0\,}\omega_{\,}\frac{k_z v_0(\alpha\omega\mp\omega_c)}{ik_zv_0+\omega\mp\omega_c}=0
\label{kone}
\,.
\end{equation*}
From which one obtains for  $n_0\ll1$, by expanding about
$\omega = k_zv_A + i \gamma +\delta \omega_r$, 
the perturbed frequency
\begin{eqnarray}
\gamma&=&-\frac{n_0}{2}\frac{k_z v_0 (\alpha k_zv_A \mp \omega_c)(k_zv_A \mp\omega_c)}{k_z^2v_0^2 + (k_zv_A \mp\omega_c)^2}
\label{gakaone}
\\
\delta\omega_r&=&-\frac{n_0}{2} \frac{k_z^2 v^2_0 (\alpha k_zv_A \mp \omega_c)}{k_z^2v_0^2 + (k_zv_A \mp\omega_c)^2}.
\end{eqnarray}
From \eqref{gakaone} we see explicitly the spurious crossover to instability at $\omega_c$ observed in Fig.~\ref{fig:gamkNH}  that occurs for $\alpha=0$.
 Similar, although progressively more complicated, formulae exist for  higher values of $\kappa$ (see, e.g., relation (117) in \cite{Podesta}), but we will not present these here.

Finally, we further explore the differences between  the Hamiltonian and non-Hamiltonian models for arbitrary isotropic equilibria by examining  the so-called small-$\gamma$ approximation for each.  Thus, we assume the resonant denominator of \eqref{disprel}  gives rise to weak damping,   and write   $\omega=\omega_r+i\gamma$, $D=D^\pm_r + i D^\pm_i$,  and then expand as usual to obtain 
\begin{equation}\label{weakdamping}
D^\pm_r(\omega_r,k_z)=0
\,,\qquad
\gamma=-\frac{D^\pm_i(\omega_r,k_z)}{{\partial}D^\pm_r(\omega_r,k_z)/{\partial \omega_r}}\,.
\end{equation}
For  $n_0\ll1$, $\partial D_r^{\pm}/\partial \omega_{r}\approx2 k_{z}v_A$,  and,  thus,  $\gamma\approx
-D^\pm_i(kv_A,k_z)/(2 kv_A)$. Using  the Plemelj relations we obtain the following form \eqref{disprel}:
\begin{eqnarray}
&&D_i^{\pm}(\omega_r,k_z)=
\label{Di}
\\
&&\hspace{1.0 cm}  
\pi n_0(\alpha\omega_r \mp\omega_c)
(\omega_r\mp\omega_c) \, \frac{ \omega_r}{k_z} F\!\left(\frac{\omega_r \pm \omega_c}{k_z}\right)\,. 
\nonumber
\end{eqnarray}
For the \underline{Hamiltonian PCS}, $\alpha=1$ and 
\begin{equation}
 D_i^\pm= \pi n_0
(\omega_r\mp\omega_c)^2 \, \frac{\omega_r}{k_z} F\!\left(\frac{\omega_r \pm \omega_c}{k_z}\right)\,. 
\label{al1}
\end{equation}
which indicates damping for both polarizations, except for the upper sign at  $\omega_r=k_zv_A=\omega_c$ where the damping vanishes. 
However, upon setting $\alpha=0$ we obtain for the \underline{non-Hamiltonian PCS}, the following:
\begin{equation}
D_i^{\pm}(\omega_r,k_z)= \pi n_0  \omega_c 
(\omega_c\mp \omega_r) \, \frac{\omega_r}{k} F\!\left(\frac{\omega_r \pm \omega_c}{k_z}\right)\,,
\label{al0} 
\end{equation}
which reveals  the  strange nonphysical crossover to instability for one of the polarizations when  $\omega_r> \omega_c$. 

For the record,  a calculation similar to that for the PCS gives for the  CCS  the dispersion relation
\begin{eqnarray}
D^{\pm}(k_z,\omega)&=& \omega^2-k_z^2v_A^2 
\\
&& 
+\,\omega\omega_c n_0 \left(\omega_c\int_{-\infty}^{+\infty}\!\frac{{F}}{k_z v_z-\omega\pm\omega_c}\,\de v_z\mp 1\right)
\nonumber
\end{eqnarray}
whence we obtain for the \underline{Hamiltonian CCS}, the following
\begin{equation}
D_i^{\pm}(\omega_r,k_z)=  \pi n_0  \omega_c^2 
 \, \frac{ \omega_r }{k_z} F\!\left(\frac{\omega_r \pm \omega_c}{k_z}\right)\,.
\label{Dpcs} 
\end{equation}
Although \eqref{Dpcs} indicates a damping rate that is different from that of the Hamiltonian PCS, it does not possess the spurious instability possessed by the non-Hamiltonian PCS.  

The damping rates indicated by \eqref{al1}, \eqref{al0}, and  \eqref{Dpcs}  have several features in common. 
First, for low frequencies, $\omega_r\ll\omega_c$,  their intended regime, they all agree. Next, they all scale with $F$ (as opposed to its derivative) which is appropriate for parallel propagating transverse waves for {\it all} isotropic equilibrium distribution functions (not just Maxwellians) \cite{KrallTravelpiece}.  For higher frequencies,   \eqref{al1}, \eqref{al0}, and  \eqref{Dpcs}   disagree so it is useful to compare with a full kinetic theory with electrons, ions, and  hot particle components.  For cold electron and ion temperatures, only the hot species contributes to the damping, and it is an elementary exercise to show that $D_i$ for this case behaves precisely as  \eqref{Dpcs}, the result for the CCS.    Thus, in this frequency range the CCS gives the best answer, although the Hamiltonian PCS may be reasonable.  Clearly, the non-Hamiltonian result is unsatisfactory.


\section{Disturbances  with $k_z=0$} 
\label{trans}

This Section presents the dispersion relation for certain linear waves propagating transversely to the magnetic field. These modes are allowed by the Hamiltonian PCS model \eqref{PCS-FuPark2}-\eqref{hybridMHD3} (with $\nabla\cdot\bU=0$), while they are forbidden by the (incompressible) non-Hamiltonian variant \eqref{PCS-FuPark}-\eqref{cc-hybrid-end}. In particular, we study the special case $\widetilde{\boldsymbol{U}}_{1\perp}=0$, which is consistent with  the incompressibility relation $\bk\cdot\tU=0$. 

In order to find the dispersion relation, we  specialize {Eq.} \eqref{velocityrelation} by setting $k_z=0$. In turn, this affects  the Vlasov perturbation \eqref{Vlasovsolution}. Since $\tU=\widetilde{U}_{1z\,}\z$ and
\[
\bB_0\times\mathcal{R}^T(\tau)\tU=\widetilde{U}_{1z\,}(\bB_0\times\mathcal{R}^T(\tau)\z)=\widetilde{U}_{1z\,}(\bB_0\times\z)=0
\,,
\]
the Vlasov perturbation \eqref{Vlasovsolution} becomes
\begin{align*}
\tf
=&\ i\alpha\int_{-\infty}^0\!\!(\bk\cdot\mathcal{R}(\tau)\bv)\left(\frac{\partial f_0}{\partial \bv}\cdot\mathcal{R}^T(\tau)\tU\right) 
e^{i\left(\bk\cdot\bX-\omega\tau\right)}\,\de\tau
\\
=&\ i\alpha\widetilde{U}_{1z}\,\frac{\partial f_0}{\partial v_z}\int_{-\infty}^0\! \bv_\perp\cdot\mathcal{R}^T(\tau)\bk\,e^{i\left(\bk_\perp\cdot\bX_\perp-\omega\tau\right)}\,\de\tau\,,
\end{align*}
which shows how the non-Hamiltonian model ($\alpha=0$) precludes the existence of transversal modes such that $\widetilde{\boldsymbol{U}}_{1\perp}=0$. In what follows, we consider the Hamiltonian case by setting $\alpha=1$.

Notice that, since $\bX_\perp$ does not depend on $v_z$, the above expression yields $\widetilde{n}_1=\int\!\tf\,\de^3v=0$. Therefore, the relation $\widetilde{n}_1=\bk\cdot\tK$ allows the case $\tK=\widetilde{K}_{1z}\z$, where
\begin{align*}
\widetilde{K}_{1 z}
=&\
i\alpha\widetilde{U}_{1z}\iint v_z\frac{\partial f_0}{\partial v_z}\int_{-\infty}^0 \bv_\perp\cdot\mathcal{R}^T(\tau)\bk\,e^{i\left(\bk_\perp\cdot\bX_\perp-\omega\tau\right)}
\\
&\hspace{3.5cm} \times\de\tau\, {\de\bv_\perp}\,\de v_z
\\
=&\
-i\alpha\widetilde{U}_{1z}\,\bk_\perp\cdot\!\iiint_{-\infty}^0\!\! \!{f}_0\,\mathcal{R}(\tau)\bv\,
 e^{i\left(\bk_\perp\cdot\bX_\perp-\omega\tau\right)}\,\de\tau\,\de^3v\,.
\end{align*}
Then, combined with  the {velocity relation  \eqref{velocityrelation} and making use of the moment equation for $\bK_1$ in Section \ref{Sec:moments},}
the special case $\widetilde{\mathbf{K}}_{1\perp}=0$ yields
$\widetilde{U}_{1z}=\widetilde{K}_{1z}$ along with the dispersion relation
\[
1+i\!\iint_{-\infty}^0 \left(\bk_\perp\cdot\mathcal{R}(\tau)\bv\right){f}_{0\,}e^{i\left(\bk_\perp\cdot\bX_\perp-\omega\tau\right)}\,\de\tau\,\de^3v=0\,.
\]
This is  an  expected Bessel function type of dispersion relation and its detailed study is left for future work.

\section{Summary and Conclusions}
\label{conclu}

After a review of hybrid kinetic-MHD models, we  presented a comparative  study of Hamiltonian and non-Hamiltonian pressure-coupling schemes, where the latter suffer by not conserving energy exactly.  In particular, the two models were compared from the point of view of linear stability and their dispersion relations were presented and  analyzed.  The special cases of pure  parallel and  perpendicular  wave propagation were considered.

Upon considering  $\kappa$ equilibria for the hot component, it was  shown that  the non-Hamiltonian PCS possesses an  instability  absent in its Hamiltonian variant and in the CCS, which is also Hamiltonian. We argued that the  instability emerging in the non-Hamiltonian model is not physically viable.  Extensive investigation of the dispersion relation will be  considered in future work.

  Although the unstable mode is of large frequency and thus outside the original intent of the PCS models, which were developed to describe low frequency behavior,   their presence would suggest results obtained from non-Hamiltonian PCS models extended into this regime should be viewed with caution.   Even if some artifice, numerical or other, were used to suppress the unphysical linear instability,  nonlinear coupling could give rise to differences in their turbulent transport behavior.

\acknowledgments
C.T.\  is indebted with J.~Carrillo, P.~Degond,  D.D.~Holm, G.~Lapenta and C.~Sovinec for several interesting discussions.  Partial support by the Institute of Mathematics and its Applications Grant \# SGS27/13 is greatly acknowledged.  P.J.M.\   gratefully  acknowledges stimulating conversations with  B.~Breizman, C.-Z.~Cheng, C.W.~Horton, F.~Pegoraro, and F.~Waelbroeck.  His research was supported by U.S.\  Dept.\ of Energy Contract \# DE-FG05-80ET-53088. {E.T.\ acknowledges financial support from the Agence Nationale de la Recherche (ANR GYPSI) and from the CNRS PEPS project GEOPLASMA.}

\appendix


\section{Poisson brackets for Hamiltonian hybrid MHD models}

\label{app1}

A Hamiltonian system is a dynamical system generated by  a  given  Hamiltonian (total energy) and a Poisson bracket in the form $\partial \Psi/\partial t= \{\Psi,H\}$, where $\Psi$ denotes the set of dynamical variables. The Poisson bracket  
$\{\cdot,\cdot\}$ must be a  bilinear, antisymmetric operator defined on the space of  function(al)s.  In addition it must satisfy  he Leibniz property
\[
\{FG,H\}=G\{F,H\}+F\{G,H\}
\]
as well as the Jacobi identity
\[
\{\{F,G\},H\}+\{\{H,F\},G\}+\{\{G,H\},F\}=0
\,.
\]
The Leibniz property,  bilinearity, and antisymmetry are easily built into the generic form of the Poisson bracket,  but  the proof of the Jacobi identity may require some effort. (See \cite{Morrison3,Morrison2005,morrison98,Morrison2} for review and the Appendix of Ref.~\cite{morrison13} for a particularly onerous  direct proof.)  Such Poisson brackets need not have the canonical form of conventional field theories and may possess degeneracy --  because of this they were called noncanonical in {Ref.}\cite{MorrisonGreene}. 

The  Poisson brackets for the Hamiltonian models of the present paper were given in  \cite{Tronci}, where it was also shown how they may be used to formulate new hybrid MHD models that conserve energy exactly.  Indeed, while exact conservation of \eqref{Ham-preMHD}  is guaranteed for the CCS model \eqref{cc-hybrid-momentum}-\eqref{cc-hybrid-end} by its noncanonical Poisson bracket  
\begin{align}
\label{PB-current-hybridMHD}
&\{F,G\}_{CCS}= \int{\bm}\cdot \left[\frac{\delta
F}{\delta{\bm}},\frac{\delta G}{\delta
{\bm}}\right]\de^3\bq
\\
& \hspace{1 cm}
- \int \rho \left(\frac{\delta F}{\delta \bm}\cdot\nabla\frac{\delta G}{\delta \rho} 
-\frac{\delta G}{\delta \bm}\cdot \nabla\frac{\delta F}{\delta \rho}\right)\de^3\bq
\nonumber\\
&\hspace{.5 cm}+
q_h
\int\! f \,\bB\cdot\bigg(\frac{\delta F}{\delta \bm}\times
\frac{\delta G}{\delta \bm} 
\nonumber\\
& \hspace{1 cm}
-\frac{\delta F}{\delta \bm}
\times
\frac{\partial}{\partial \bp}\frac{\delta
G}{\delta f}+\frac{\delta G}{\delta \bm}
\times
\frac{\partial}{\partial \bp}\frac{\delta
F}{\delta f}\bigg)\dvol
\nonumber\\
&\hspace{.5 cm}+
\int \!f\bigg(\left\{\frac{\delta
F}{\delta f},\frac{\delta G}{\delta f}\right\} 
\nonumber\\
&\hspace{1 cm}+q_h\,\bB\cdot\frac{\partial}{\partial \bp}\frac{\delta
F}{\delta f}
\times
\frac{\partial}{\partial \bp}\frac{\delta
G}{\delta f}\bigg)\dvol
\nonumber\\
&
+
\int \bB\cdot\left(\frac{\delta F}{\delta \bm}\times\nabla\times\frac{\delta G}{\delta \bB}-\frac{\delta G}{\delta \bm}\times\nabla\times\frac{\delta F}{\delta \bB}\right)\de^3\bq
\,,
\nonumber
\end{align}
the PCS models available in the literature fail to conserve energy exactly. For the Hamiltonian PCS (HPCS)  \eqref{PCS-FuPark2}-\eqref{hybridMHD3}, exact conservation of \eqref{cc-hybrid-momentum}-\eqref{cc-hybrid-end} follows {from the} Poisson bracket 
\begin{align}
&\{F,G\}_{\tiny HPCS}= \int{\bM}\cdot \left[\frac{\delta
F}{\delta{\bM}},\frac{\delta G}{\delta
{\bM}}\right]\de^3\bq
 \label{PB-pressure-hybridMHD}
 \\
&
\hspace{.5cm} -\int \rho \left(\frac{\delta F}{\delta \bM}\cdot\nabla\frac{\delta G}{\delta \rho} 
-\frac{\delta G}{\delta \bM}\cdot \nabla\frac{\delta F}{\delta \rho}\right)\de^3\bq
\nonumber
\\
&
\hspace{.5cm}+ \int f\left(\bigg\{\frac{\delta
F}{\delta f},\frac{\delta G}{\delta f}\right\}
\nonumber\\
&
\hspace{1.5 cm}
 +q_h\, \bB\cdot\frac{\partial}{\partial \bp}\frac{\delta
F}{\delta f}
\times
\frac{\partial}{\partial \bp}\frac{\delta
G}{\delta f}
\bigg)\dvol
\nonumber\\
&
+ \int f\left(\left\{\frac{\delta
F}{\delta f},\bp\cdot\frac{\delta G}{\delta \bM}\right\}-\left\{\frac{\delta
G}{\delta f},\bp\cdot\frac{\delta F}{\delta \bM}\right\}\right)\dvol
\nonumber\\
&
+ \int \bB\cdot\left(\frac{\delta F}{\delta \bM}\times\nabla\times\frac{\delta G}{\delta \bB}-\frac{\delta G}{\delta \bM}\times\nabla\times\frac{\delta F}{\delta \bB}\right)\de^3\bq
\,.
\nonumber
\end{align}
In the above formulas, ${[\boldsymbol{X},\boldsymbol{Y}]}:=-{(\boldsymbol{X}\cdot\nabla)
\boldsymbol{Y}}+{(\boldsymbol{Y}\cdot\nabla) \boldsymbol{X}}$ is
minus the commutator on vector fields.
The proof that the above bilinear, antisymmetric operators are indeed  Poisson brackets (satisfying Leibniz and Jacobi) can be carried out by  explicit verification. However,  upon recognizing that these brackets are composed of terms of the original bracket of  MHD \cite{MorrisonGreene} and that of the Maxwell-Vlasov system \cite{Morrison2bis, Morrison3,MaWe1,MaWeRaScSp}, together with later work on the two-fluid system \cite{Spencer,SpKa}, it is not difficult to ascertain the validity of the Jacobi identity.  

Alternatively, one can begin with an action principle and derive the Poisson brackets, thereby ensuring the Jacobi identity.   Such a  Lagrangian formulation of the PCS Eqs.~\eqref{PCS-FuPark2}-\eqref{hybridMHD3} was given in \cite{HoTr2011}.  We remark also that an action principle derivation of a linearized PCS model was presented in \cite{Bri94}.  
%


\section{Derivation of dispersion relation  for $\mathbf{k}_{\perp}=0$}
\label{sec:disprel}

This appendix contains the main steps leading to the dispersion relation \eqref{disprel}. The starting point is the observation that   $\widetilde{U}_{1 z}=({\mathcal{A}^T\tU})_z = ({\bB_0\times\mathcal{A}^T\tU})_z=0$ forces relation  \eqref{pre-disprel} to possess only planar components. Then, one can write the dispersion relation as
\ben
&&\left(\frac{\omega^2-k_z^2 v_A^2}{\omega}-\int\!k_z\, \frac{\partial\bar{f_0}}{\partial v_z}\left(i\alpha k_z v_z\mathcal{A}_{11}+\beta v_A\mathcal{A}_{12}\right)\de v_z\right)^2 
\nonumber\\
&&\hspace{1.5 cm}=
\left(\int\!k_z\, \frac{\partial\bar{f_0}}{\partial v_z}\left(i\alpha k_z v_z\mathcal{A}_{12}+\beta v_A\mathcal{A}_{11}\right)\de v_z\right)^2\,, 
\nonumber
\een
where we recall the definitions \eqref{A11}-\eqref{A12}. (Notice that $\bar{f_0}$ denotes the distribution function divided by the constant bulk particle density).  
Then, after some   computations and upon restoring  physical constants, one is led to
\ben
&&\frac{\omega^2-k_z^2 v_A^2}{\omega}+\alpha\int_{-\infty}^{\infty} \frac{(k_zv_z)^2\bar{f}_0}{k_z v_z-\omega\pm\omega_c}\,\de v_z
\label{diprel}\\
&&\hspace{1.8cm}=\pm
 \beta \omega_c\int_{-\infty}^{\infty}\frac{k_zv_z\bar{f_0}}{k_z v_z-\omega\pm\omega_c}\,\de v_z
 \,.
\nonumber
\een
The integrals of (\ref{diprel}) are then rearranged  as follows:
\begin{align*}
\int_{-\infty}^{\infty}\frac{k_z^2v_z^2\bar{f_0}}{k_z v_z-\omega\pm\omega_c}\,\de v_z&
\\
& \hspace{-3 cm}=
(\omega\mp\omega_c)\left(n_0+(\omega\mp\omega_c)\int_{-\infty}^{\infty}\!\frac{\bar{f_0}}{k_z v_z-\omega\pm\omega_c}\,\de v_z\right)
\\
\int_{-\infty}^{\infty}\frac{\omega_c k_zv_z\bar{f_0}}{k_z v_z-\omega\pm\omega_c}\,\de v_z&
\\
& \hspace{-3 cm}=
\omega_c\left(n_0 
+(\omega\mp\omega_c)\int_{-\infty}^{\infty}\!\frac{\bar{f_0}}{k_z v_z-\omega\pm\omega_c}\,\de v_z\right)
\,.
\end{align*}
Finally, upon recalling that $\beta=1-\alpha$, we  write the dispersion relation as
\begin{multline*}
\frac{\omega^2-k_z^2 v_A^2}{\omega}= 
\big(\pm\,
\omega_c (1-\alpha) - \alpha(\omega\mp\omega_c)\big)
\\
\hspace{.5cm} \times
\left(n_0+(\omega\mp\omega_c)\int_{-\infty}^{\infty}\!\frac{\bar{f_0}}{k_z v_z-\omega\pm\omega_c}\,\de v_z\right)\,, 
\end{multline*}
which eventually reduces to \eqref{disprel}.

\end{document}